\documentclass{llncs}
\usepackage{eurosym}
\usepackage{graphicx}
\usepackage{verbatim} 
\usepackage{enumitem}
\usepackage[cp1252]{inputenc}
\usepackage{psfig}
\usepackage[T1]{fontenc}
\usepackage{textcomp}
\usepackage{lmodern}
\usepackage{cite}

\renewcommand{\contentsname}{Contenidos}

\newcommand{\codigo}{\tt\small}

\begin{document}

\title{Extending ChatGPT with a Browserless System 
for Web Product Price Extraction}

\author{Jorge Lloret-Gazo}
\authorrunning{Jorge Lloret}   
%
\tocauthor{Jorge
Lloret-Gazo(Universidad de Zaragoza)}

\institute{Dpto. de Inform\'atica e Ingenier\'{\i}a de Sistemas.\\
Facultad de Ciencias. Edificio de Matem\'aticas.\\ Universidad de
Zaragoza. 50009 Zaragoza. Spain.\\ \email{jlloret@unizar.es}}
\maketitle   
\begin{abstract}
With the advenement of ChatGPT, we can find very clean, precise answers to a varied amount of questions. 
However, for questions such as 'find the price of the lemon cake at zingerman's', 
the answer looks like `I can’t browse the web right now'.
In this paper, we propose a system, called Wextractor, which extends ChatGPT to answer questions as the one mentioned before.
Obviously, our system cannot be labeled as `artificial intelligence'. Simply, it offers to cover a kind of transactional search that is not
included in the current version of ChatGPT.
Moreover, Wextractor includes two improvements with respect to the initial version: social extraction and 
pointing pattern extraction to improve the answer speed.
\end{abstract}

\section{Introduction}

Since its launch, ChatGPT has grown quickly to become a well-known question-and-answer system.
Numerous exam-related use cases involving ChatGPT have had their performance examined, 
with different degrees of scientific rigor ranging from in-depth research to anecdotal evidence. 
Use cases include excellent success on the United States Medical Licensing Examination~\cite{kung2023performance}. 
For these and other reasons, large language models (LLMs) are expected to have an effect on a variety of fields and 
be used as assistants by a variety of professionals .

However, there are many easier questions that ChatGPT is not able to answer, 
for example, those related to queries that requires browser capabilities and, in particular,
with product prices on the web.

On the other hand, businesses are interested in extracting product prices 
from different web sites and in following them up
during a certain period of time for different purposes. 
For example, an e-commerce enterprise is interested in the price of pairs of shoes
listed on different competing websites with the purpose of adapting their prices when
competence prices fall below a certain threshold.

In this paper, we propose the combination of ChatGPT with an enhanced version of
Wextractor~\cite{lloret2020browserless} to answer questions about product prices.

To be specific, in paper~\cite{lloret2020browserless}, 
we proposed an architecture for extracting prices that 
applies the following
four steps: (0) browserless extraction of raw HTML, (1) fragmentation,
(2) rule application and (3) price extraction.
In step (0), raw HTML is obtained from the url of the page.
In step (1), fragments that contain clues
of the prices are found. In the next step, rules previously designed
by a rule designer
are applied to the fragments.
As a result, several fragments, that contain several candidate prices, remain and
one of them is chosen as the target value.
Step 0 is based on the browserless concept of~\cite{Fayzrakhmanov:2018:BWD:3178876.3186008}.
The first step is inspired by the segmentation of~\cite{DBLP:dblp_journals/tkde/ZhaiL06}
while, to the best of our knowledge, there is no proposal of easy-to-specify rules in the
literature for the purpose of extracting prices.

In this paper, we extend the original architecture with two ideas: (a) social extraction and
(b) pointing pattern extraction. The first idea consists of storing prices of objects coming from searches of other users
and using these prices to answer subsequent queries that occur near the initial
search. The second idea is to use automatic pointing
pattern extraction. A pointing pattern is a regular expression that points to where the
entity price is. 

Then, the second and subsequent times an extraction is performed on a page,
first, a social search will be performed. If the price is available in the social storage, it
will be returned as a result. Otherwise, 
the pointing pattern of the page is used. If this extraction does not find any price,
the process begins and steps (1)-(3) are applied again (step (0) was previously applied,
so it is not applied again).

Taking into account our enhanced architecture, questions on product prices 
can be dispatched from ChatGPT to Wextractor, extending
in this way the capabilities of ChatGPT.

The contribution of this paper is threefold:
\begin{enumerate}

\item{We offer an answer to ChatGPT-uncovered questions}

\item{We incorporate a social module into Wextractor in which people act 
as sensors that transmit information from the market to our architecture.
Therefore, some searches can be answered with 
the information previously gathered from users. This avoids computationally expensive queries.}

\item{Finally, we incorporate a pointing pattern module so that a 
faster search can be performed using a pointing pattern.}

\end{enumerate}

The rest of the paper is organised as follows. Section 2 is devoted
to defining the problem and the challenges. In Section
3, we explain the architecture and in Section 5, an implementation
of it. Section 4 is devoted to the connexion between
ChatGPT and Wextractor. In the last sections, 
we deal with the evaluation, related work,
conclusions and future work.

\section{Problem definition and description of the architecture}

\begin{figure*}[t]
\centerline{\includegraphics[width=80mm]{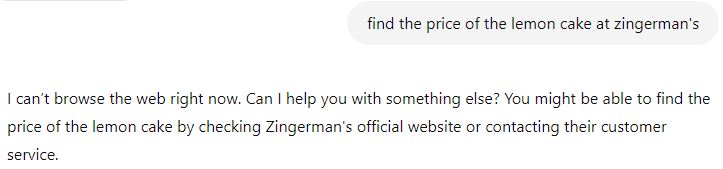}}
\caption{Answer of ChatGPT to a query about Zingerman's cake}
\label{fig:answer}
\end{figure*}

Although ChatGPT is a powerful tool, there are questions 
which remain not satisfactorily answered.
For example, let us consider a query
such as `price of lemon cake at Zingerman''s'.
The answer of ChatGPT is shown in Figure~\ref{fig:answer}.
That is, ChatGPT cannot deal with real-time queries or with queries that need
browser capabilities.

On the other hand, our Wextractor architecture responds
to questions as follows: 
Given a detail web page {\tt\small pg} of an entity {\tt\small e} with url {\tt\small u}, 
extract the price of {\tt\small e} from page {\tt\small pg}. 

An example of this problem is query (q1): Given the url

\begin{center} 
{\codigo https://www.zingermans.com/Product.aspx?ProductID=A-ZDV}
\end{center}

find the price of the lemon cake.

\begin{figure*}[t]
\centerline{\includegraphics[width=60mm]{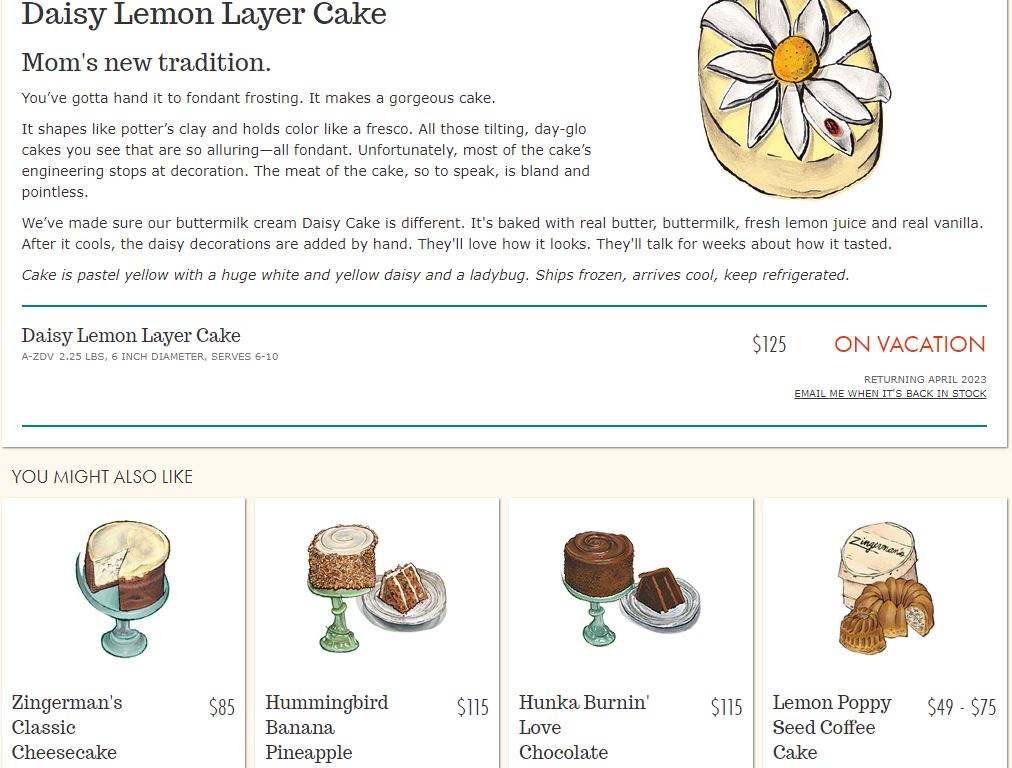}}
\caption{Fragment of the Zingerman web page as of March 2024}
\label{fig:interface1}
\end{figure*}

The problems we solve in this paper are:

\begin{enumerate}
\item{How to improve the answer speed of Wextractor (see Section 6).}
\item{How to connect ChatGPT with Wextractor (see Section 4)
to answer real-time queries on prices with ChatGPT.}
\end{enumerate}

\begin{figure*}
\centerline{\includegraphics[width=120mm]{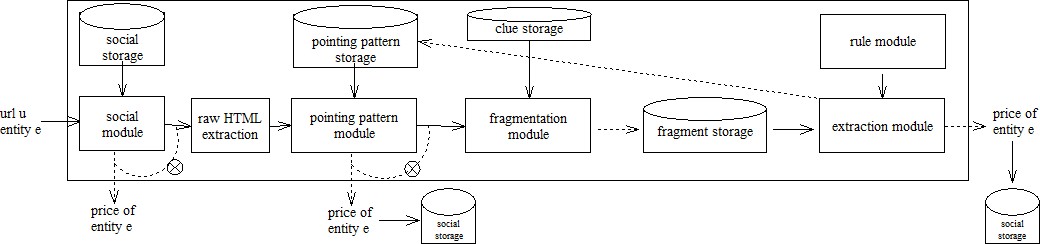}}
\caption{Our architecture}
\label{fig:architecture}
\end{figure*}

The solution of the first problem is expressed 
in the following four steps. As an example, we will use query (q1).
Togehter with the steps, we summarize our architecture extended with the social and the pointing pattern
modules (see~\ref{fig:architecture}). Additional details of the original architecture can be found on~\cite{lloret2020browserless}.

{\bf Step 0. Social extraction} 

This extraction is done by the {\em social module}. 
It receives the url u of a detail page for an entity e and searches
the price associated with the url u in the {\em social storage}. If the price is found,
it is returned provided that the difference of the actual time with the time of the price in the social storage falls within a limit.
If the URL  is not available or the limit is overcome, then jump to the step 1.

This module is inspired by Waze~\cite{vasserman2015implementing}. In its page, we can read 
 ``Nothing can beat people working together''. The idea behind this module is that
people act as sensors that transmit the information from the market to the Wextractor architecture.
This information gathered by the sensors is elaborated inside our architecture and offered curated
to other people doing similar searches through ChatGPT.

In this way, information is bidirectional. 
As far as I know, the information flows from the buyer to the seller and 
remains stored there or is stored in Google.
According to our idea, other buyers can directly access the information
of other buyers.

{\bf Step 1. Browserless extraction of raw HTML}  

From the url u of the page, the raw HTML code is extracted without interaction from a browser.

{\bf Step 2. Pointing pattern extraction} 

This extraction occurs when the social extraction does not give any results.
It needs the following two modules:

{\em Pointing pattern storage} This component stores pairs formed by the url of a page and a pointing
pattern, i.e., a regular expression that  points to the price of the page.

{\em Pointing pattern module} This module receives the url of the page of an entity and 
searches in the {\em pointing pattern storage} for any 
pointing pattern associated with the url u. If a pointing pattern is found, 
the price is extracted from the raw HTML of the page obtained in Step 1.

As a result, the price of the entity or nothing if the pointing pattern does not work is obtained. In this last case or if the page with url u
was not previously accessed, an extraction from scratch begins
in the next step.

{\bf Step 3. From scratch extraction}  
From this point on, a from scratch extraction is performed. This extraction
occurs in one of the following cases:

\begin{enumerate}
\item{it is the first time a page is accessed}
\item{the page was previously accessed but neither the social extraction nor the pointing pattern extraction gave any results}
\end{enumerate}

It needs the following modules:

{\em Clue storage} This stores strings representing the currencies of interest.
For example, EUR for the currency euro.

{\em Fragmentation module} This module extracts, from the raw HTML of a page with url u, the
fragments that contain possible prices of entity e. To do so, it uses
the {\em clue storage} to identify the places where prices are.
Examples of clues are \&euro; or \$.
For query (q1), 149 fragments were extracted, two of them being:

(f1) {\tt\small <span id="ctl00" class="price">\$ 125</span>}

(f2) {\tt\small <div class="saving">SAVE10\%=\&euro;12.80</div>}

All these fragments are stored in the {\em fragment storage}.

{\em Rule module} The rule module is composed of an ordered set of Condition-Action rules elaborated by the designer.
We will use {\em discarding rules} to discard fragments verifying conditions 
that indicate the absence
of the target price. Then, a weight of one is added to these fragments 
and zero to the rest of the fragments.
In the end, fragments with a total weight equal to zero are selected.

In the example of the cake, the discarding rules search for properties
of the fragments that indicate that they do not contain the price of the cake.
An example of a discarding rule, called {\tt\small semr1}, is
`discard those fragments that contain the word SAVE'.
(f2) is an example of a fragment discarded by the {\tt\small semr1} discarding rule.

{\em Extraction module} This module applies the rules of the {\em rule module} to the fragments of the
{\em fragment storage} and determines the right price for entity e. 
This module works as follows. First, the fragments of the page are loaded from the {\em fragment storage} and their
weight is initialized to zero. Next, the rules are applied to the fragments 
updating the weights accordingly. Then, the number of candidate prices is calculated
where the candidate price are those whose fragments have weight equal to zero.
If only one candidate price is found this is the target price. 

In the example of the cake, fragment (f1) is the only non-discarded fragment and from it the
value \$125 is obtained. 

However, no price or more than one
price could also be retrieved. To remedy this, we have defined recovering rules, i.e., rules that try to rescue the correct
price from previously discarded prices. After applying these rules, 
if only one  price is found this is the target price. Otherwise, there is no solution.

If  the user wants to check the price of an entity again, we have two options. 
One of them is to repeat a from scratch extraction. 
The second one is to try to use the information of the from scratch extraction to speed up the process.
Because our method relies on detecting HTML fragments and, in general, the web page structure 
does not change too much with time,  we have chosen the second option.

Following the second option, the non-discarded fragments 
of the from scratch extraction are used to build
a pointing pattern, that is, a regular expression that matches the price.
For example, after applying the from scratch extraction in query (q1), 
only fragment (f1) remains. From this fragment,
the pointing pattern pp1 {\codigo id="ct100" class="price">\$[0-9]\{2,3\}$\backslash$.[0-9]\{1,2\}} is built for subsequent extractions. 
It is worth noting that when a new extraction is performed with this pattern, the price of 125\$ could have fallen below 100\$. 
For this reason, the first part of the numeric pattern is {\codigo [0-9]\{2,3\}}, that is, prices of less than 100\$ are also matched by the pattern.

According to the previous discussion, two additional actions are taken after the from scratch extraction success:

\begin{enumerate}
\item{A pointing pattern is extracted and stored in the {\em pointing pattern storage} to be used in future searches}
\item{Quaterns (e,u, p, t) are stored in the {\em social storage} where e is an entity, 
u is the url of a detail page for entity e in page u and t is the timestamp where the price was found}
\end{enumerate}

Then, the next time ChatGPT demands the price of an entity e, it will dispatch the query to the enhanced
Wextractor architecture which, in turn, will offer an answer, such as the url of the page together with the price of 125 \$. 
Otherwise, the query would remain without an answer inside ChatGPT.

\section{Combined architecture with ChatGPT and Wextractor}

In this Section, we explain how to combine the potential of ChatGPT and
Wextractor to answer price questions.

\begin{figure*}[t]
\centerline{\includegraphics[width=80mm]{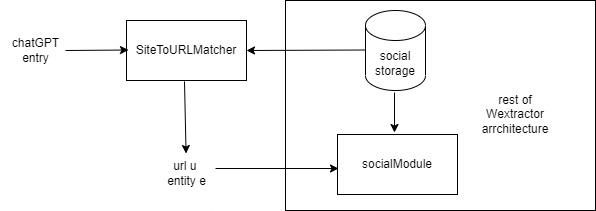}}
\caption{Connection between ChatGPT and Wextractor}
\label{fig:extension}
\end{figure*}

The way of working is as follows (see Figure~\ref{fig:extension}): The user issues a query 
against ChatGPT which,
 extracts the tuple (e0,s0) from the query, where e0 is an entity for which the price
must be found in site s0. For example, from the query `find the price 
of the lemon cake at Zingerman', ChatGPT extracts the pair
(`lemon cake','Zingerman').

On the other hand, the Wextractor architecture is fed with user searches for
prices. In these searches, the user provides a url from which a price is obtained.
However, this is not the case for ChatGPT where the user provides a textual query,
without URLs, such as `find the price of
of the lemon cake at Zingerman'.
So, in order to communicate ChatGPT and Wextractor, we need a new module to
find the urls of Wextractor that correspond to the site of the ChatGPT query.

This new module is called {\codigo siteTOURLMatcher} and is responsible 
for finding every url in the social
storage that belongs to the site (see Figure~\ref{fig:extension}).
As a result, the module obtains a pair (possibly empty) (u,p) where 
 u is an url corresponding to the site s0 and p is the price
of e0 in the page with url u.

For example, for the pair (`lemon cake',`Zingerman'), the 
siteTOURLMatcher module queries the social storage of Wextractor
and finds the pair 

({\codigo https://www.zingermans.com/Product.aspx?ProductID=A-ZDV}, \$125)

So, this pair is sent back to ChatGPT as an answer to the
query `find the price 
of the lemon cake at Zingerman'.

\section{An implementation of the Architecture}
\label{implementation}
In this section, we describe a 
specific implementation of our architecture.
As designers, we have chosen to implement
our architecture by using relational databases 
and SQL, all glued with a procedural language for
relational databases. 

\begin{table}[htbp]
\caption{Pseudocode for finding the price of a detail page} 
\label{tab:main}
\begin{tabular}{p{0.25cm}p{12cm}}
& Algorithm Wextractor\\\hline
& Input: url u of the page pg of an entity e\\
& Output: 

success=-1.  The page is not available

success=0. No price was found.

success=1.  One price was found by from scratch.

success=2.  One price was found by pointing pattern extraction.

success=3.  One price was found by social extraction.

price of the entity when success=1 or 2 or 3
\\\hline
& Pseudocode\\

1.&{\codigo  \hspace{1.5mm} success<-false}\\

2.&{\codigo  \hspace{1.5mm} if isAvailablePage(u) then}\\

3.&{\codigo  \hspace{3mm} found=getSocialPrice(u, price, timestamp)}\\

4.&{\codigo  \hspace{3mm} if(found) then}\\

5.&{\codigo  \hspace{4.5mm} if(sysdate-timestamp<socialExtractionValidity) then success=3}\\

6.&{\codigo  \hspace{4.5mm} else found=false}\\

7.&{\codigo  \hspace{3mm} end if}\\

8.&{\codigo  \hspace{3mm} end if}\\

9.&{\codigo  \hspace{3mm} if (!found)}\\

10. &{\codigo  \hspace{4.5mm} found=getPointingPattern(pointingPattern, u)}\\

11.&{\codigo  \hspace{4.5mm} if (found) then}\\

12.&{\codigo  \hspace{6mm} if(sysdate-timestamp<pointingPatternValidity) then}\\

13.&{\codigo  \hspace{7.5mm} getHTMLCode(HTMLcode, u)}\\

14. &{\codigo  \hspace{7.5mm} found=doPPExtraction(HTMLCode, pointingPattern, price)}\\

15. & {\codigo  \hspace{7.5mm}if (found) then  success=2}\\

16. & {\codigo  \hspace{7.5mm}else  found=false}\\

17. & {\codigo  \hspace{7.5mm}end if}\\

18. & {\codigo  \hspace{6mm}else  found=false}\\

19.&{\codigo  \hspace{3mm} end if}\\

20. & {\codigo  \hspace{6mm}end if}\\

21.&{\codigo  \hspace{3mm} end if} \\

22.&   {\codigo  \hspace{4.5mm}if !found}\\

23.&{\codigo  \hspace{6mm} getHTMLCode(HTMLcode, u)}\\

24.&     {\codigo  \hspace{6mm} found=doFromScratchExtraction(HTMLCode,s)}\\

25. & {\codigo  \hspace{6mm}if (found) then  success=1}\\

26.&       {\codigo  \hspace{7.5mm}findAndStorePointingPattern(HTMLCode, u)}\\

27.&       {\codigo  \hspace{7.5mm}storageInSocialStorage(entity, u,price, timestamp)}\\

28.&       {\codigo  \hspace{6mm}else success=0}\\

29.&      {\codigo  \hspace{6mm}end if}\\

30.&     {\codigo  \hspace{4.5mm}end if}\\

31.& {\codigo  \hspace{1.5mm}else}\\

32.&   {\codigo  \hspace{3mm}success=-1 }\\

33.& {\codigo  \hspace{1.5mm}end if}\\

\end{tabular}

\end{table}

Table~\ref{tab:main} shows the main algorithm and
combines the three modules of the architecture. 

We have set a time of validity of a social extraction 
({\codigo socialExtractionValidity}) and a time of validity of 
a pointing pattern extraction ({\codigo pointingPatternValidity}) 
where 
{\codigo socialExtractionValidity<pointingPatternValidity}.
The 

{\codigo socialExtractionValidity} means that a correct price obtained 
at time {\codigo t0} is valid for any other request until time
{\codigo t0+socialExtractionValidity}. If this limit is overcome, the price is extracted 
by a pointing pattern search until {\codigo t0+pointingPatternValidity}.
We consider that social extraction is less durable because it deals with the prices of the object 
and pointing pattern  
is more durable because it deals with the structure of the page, which changes less frequently.

In the algorithm, a social extraction is first performed (lines 3 to 8). 
If this extraction does not give the expected results, 
a pointing pattern extraction is performed (lines 9 to 21). 
In particular, the first time an extraction is performed, 
as there are  no previously available patterns, 
a from scratch extraction is always performed (lines 22 to 30).

In detail, the algorithm is as follows:
In line 2, the existence of a page with url u is checked. 
If the page is not available, success=-1 is returned (line 32). 

{\bf Social extraction}
If the page is available in the social storage (line 3) and the time of the social storage
falls within a limit, then success=3 (line 5) and the price fetched 
from the social storage is the result.

{\bf Pointing pattern extraction}
Otherwise, in line 10, a pointing pattern is searched in the pointing pattern storage for page u.
If it is found and the time of the pointing pattern 
falls within a limit, a pointing pattern extraction is performed (line 14), using
the algorithm {\codigo doPPExtraction}.
At the end, if a price is found, it is the result and success is set to 2 (line 15). 
Otherwise, the pointing pattern extraction failed because,
for example, the structure of the page
has changed so that the pointing pattern does not find the correct price.

{\bf From scratch extraction}
If the previous extractions did not find any solution, 
a from scratch extraction is performed (line 24).
If a price is found, it is the result and success is set to 1. 
Moreover, a pointing pattern is determined (line 26) 
and added to the pointing pattern storage 
for use in subsequent pattern-based extractions.
On the other hand, 
the entity, url, found price and the
timestamp in which the price was found are stored in the social storage (line 27).
If no price is found, success is set to 0 (line 28).

\section{Experimental validation}
\label{experimental}
Our architecture could be tested in a real environment, but it would take 
several months to complete.
Instead, we prepared a simulation based on the data gathered 
from our 735-page dataset~\cite{lloret2020browserless}.
To be specific, our previous
dataset served to set the number of pages and the probability of success. 
These initial data have
been combined with our hypothesis 
that the request of pages follows a Zipf distribution. Let us dive into the details. 

{\bf Dataset}
We searched for websites where we could find very visited shopping websites 
with the following requirements:
(a) the list must be representative of the most popular shopping websites globally,
and (b) the list must consist of shopping websites in English so that we would have the means to
analyze the data collected from the websites.

Previously, we could retrieve the list from places like Alexa using the Top Sites API [9] or 
WebShrinker or 5000best.com. However, at the time of writing, Alexa is not available, 
WebShrinker does not
have an open version of the data and 5000best.com 
does not offer any information about shopping websites.
Moreover, a Google search on the topic gives very generic results.
For these reasons, we will use our original list of 735 pages published 
in the paper~\cite{lloret2020browserless}. 

{\bf Simulation Features} The general features of our simulation are as follows:

\begin{enumerate}

\item{The simulation runs on {\codigo Npages} web pages from  dataset D.}

\item{ {\codigo N>=0} distinct requests are made of dataset D 
in the integer values of the time interval $\left[1, N\right]$, that is, a request per unit of time}

\item{Each request has the form (p, t), where p belongs to dataset D 
and t is an integer time of interval $\left[1, N\right]$}

\item{The pages are demanded according to a Zipf distribution of parameters a=b=1. 
Using the algorithm~\cite{zipf}, we generated N numbers between 1 and {\codigo Npages} 
each simulating a requested page.
Without loss of generality, we supposed that the pages are ordered according to their demand. 
That is, page 1 is the most demanded and 
page {\codigo Npages} is the least demanded. Therefore, page 2 is half demanded than
page 1.}

\item{The probability of success (that is, the price was found) 
of the from scratch search on each page is  p 
and the probability of failure is 1-p}.

\end{enumerate}

{\bf Simulation at Work}
 
{\bf Exact numbers}
For simulating our architecture at work, we used the following figures:
{\tt Npages=735, N=50.000, p=579/735, 

socialExtractionValidity=10, pointingPatternValidity=20.}

For determining p,  we used the fact that our Wextractor architecture produces 
exact results for 579 out of 735 pages~\cite{lloret2020browserless}. 
Therefore, we calculated
the probability of success for each page in a from scratch extraction as p=579/735.

{\bf Simulation database} 
We implemented the simulation using the Oracle database 21g and the PL/SQL
programming language. The database upon which the simulation is built 
consists of the following tables.

{\codigo fromScratchResultOfExtraction(idPage, success)}

{\codigo zipfNumber(id, numero)}

{\codigo resultOfExperiment(idPage, success, time)}

The data input are in tables
{\codigo fromScratchResultOfExtraction} and{\codigo zipfNumber}. 
The table {\codigo fromScratchResultOfExtraction}
has been filled with 735 rows, one per page, and the success column 
is calculated according to the
probability p=579/735. The table zipfNumber 
is filled with N values of idPage where each value ranges  between 1 and {\codigo Npages}. 
The values follow
a Zipf distribution, as stated in point 4.

The output data are in table {\codigo resultOfExperiment}. This table is completed after executing the 
simulation algorithm below under
the assumption that the results of the from scratch extraction are  in table {\codigo fromScratchResultOfExtraction}.
The table{\codigo resultOfExperiment} will contain N rows and each row is composed
of the id of the queried page, the success of the search and the time of the search. 
The possible values for success are:
0 if no price is found, 
1 if the price is found from scratch, 
2 if the price is found by pointing pattern extraction and
3 if the price is found by social extraction.

{\bf Simulation algorithm} With this database, the simulation algorithm is as follows

{\small\tt
for each i in 1..N

  \hspace{2mm} page<--getZipfNumber(i)

  \hspace{2mm}  success<-- getSuccess(page, i)

  \hspace{2mm}  storeSuccess(page, success)

end for
}

{\bf Results and conclusions} 
The number of successful experiments were 43130 out of 50.000. 
This means  86.26\% success.

\begin{table}[t]%
\caption{Frequency of success}
\label{tab:frequency}
\centering
\begin{tabular}{|p{1.5cm}|p{1.5cm}|p{1.5cm}|p{1.5cm}|}
\hline
Page&Success=1 &Success=2& Success=3\\\hline
1& 412& 1382 & 5073 \\\hline

\end{tabular}
\end{table}%

With respect to individual pages, in Table~\ref{tab:frequency},  
we gather the frequency of success for the most frequent page, number 1. 
Page number 1 was requested 6867 of 50.000 times. As can be seen in
Table~\ref{tab:frequency}, in most cases it was not needed a from scratch
extraction, leading to the quickest answer by the social extraction or pointing pattern extraction. 

Finally, for those frequent pages for which our Wextractor does not find any solution, 
it would be better to manually find  a pointing
pattern. If we found manually a pointing pattern for the 10\% most frequent pages, 
the success rate would increase 
from 86.26\% to 90.67\%.

{\bf Reproducibility} Following the reproducibility recommendation of the ACM SIGMOD and of VLDB, we have made
available at~\cite{dataset} the script for an implementation of the architecture, the text files with the HTML code of
the web pages of our dataset and the code of the simulation of how the architecture works.

\section{Related Work}

Recently, ChatGPT has emerged as a question-and-answer system. 
However, as evidenced in this paper, certain inquiries, 
like web prices, remain challenging to address effectively. 
To our awareness, no existing research has extended ChatGPT to encompass the capability of extracting prices from the web, unlike our present work. 
This paper aims to bridge this gap by introducing an innovative approach to augmenting ChatGPT with price extraction capabilities. 
By doing so, we contribute to enhancing ChatGPT's utility in addressing a broader range of inquiries, particularly those pertaining to e-commerce and data mining. 
Our endeavor underscores the importance of continually advancing 
AI systems to tackle increasingly complex tasks and highlights the potential for extending ChatGPT's functionality beyond its current scope. 
Through this research, we aim to facilitate more comprehensive and insightful interactions with ChatGPT, thereby enriching its practical applications in various domains.

With respect to the problem of 
extracting data from web pages, it appeared early in the year 2000 and it is an important task in e-commerce and data mining. 
The proposed solution has essentially been the use of wrappers, that is, are programs or scripts that help extract structured data from web pages by identifying specific HTML elements or patterns. 
Reviewing the literature, we have three main ways of generating wrappers~\cite{DBLP:dblp_journals/tkde/ChangKGS06,DBLP:journals/ftdb/Sarawagi08,ferrara2014web}:
manual, wrapper induction, visual based.

{\bf Manual creation. }
The manual creation of wrappers~\cite{DBLP:dblp_conf/iiwas/StarkaHN13} is a traditional approach used for extracting data from web pages. It involves manually designing and coding scripts or programs 
to extract specific information from HTML documents. 
This method requires a deep understanding of web page structures and the ability to write custom code to navigate and extract data accurately.
In the manual creation process, developers typically analyze the target web pages, identify the relevant HTML elements, and define the extraction rules.
These rules can include techniques such as pattern matching, XPath queries, or CSS selectors to locate and extract the desired data, such as prices. 
The extracted data is then processed and transformed for further analysis or integration into other systems.

One advantage of manual wrapper creation is its flexibility and control over the extraction process. 
Developers can customize the extraction logic to handle various website layouts and adapt to changes over time. It also allows for fine-grained control over data quality and validation.
However, manual wrapper creation can be time-consuming and labor-intensive, especially when dealing with a large number of websites or complex web page structures. 
It requires technical expertise and ongoing maintenance to keep the wrappers up-to-date as websites evolve.

To streamline the process, researchers have explored semi-automatic or automatic methods for wrapper creation, 
such as machine learning algorithms or rule induction techniques. These approaches aim to reduce the manual effort required 
by automatically learning extraction rules from training data or by inferring patterns from example pages.

{\bf Wrapper induction creation}
Wrapper induction~\cite{DBLP:dblp_conf/ijcai/KushmerickWD97,DBLP:dblp_journals/ml/Soderland99,DBLP:dblp_journals/dke/LaenderRS02}, 
also known as automatic wrapper generation, is an approach that aims to automate the process of creating wrappers 
for data extraction from web pages. Instead of manually designing and coding extraction rules, wrapper induction techniques use machine learning algorithms 
to automatically learn the extraction patterns from example web pages.

In the wrapper induction process, a set of example web pages with the desired data is provided as training data. 
The machine learning algorithm analyzes the HTML structures, tags, attributes, and textual content of these pages to identify patterns and create extraction rules. 
These rules are then used to extract the desired data from unseen web pages that have a similar structure.

One of the key advantages of wrapper induction is its ability to handle a large number of web pages and adapt to changes in website layouts. 
As the algorithm learns from the training examples, it generalizes the extraction rules to handle variations in web page structures 
and accurately extract the desired data. This makes it a scalable and efficient approach for data extraction from diverse sources.

Wrapper induction techniques can employ various machine learning algorithms, such as decision trees, rule-based systems, or sequence models, 
depending on the characteristics of the data and the extraction task. Some approaches also incorporate active learning, 
where the algorithm iteratively selects informative examples for labeling by a human expert, further improving the accuracy of the wrapper generation process.

However, wrapper induction may face challenges when dealing with noisy or complex web pages that deviate 
from the patterns observed in the trabfining data. Handling dynamic content, JavaScript-driven interactions, or dynamically generated elements 
can also be challenging for wrapper induction algorithms.

{\bf Visual based wrapper creation}
Visual-based wrapper creation (Mozenda~\cite{Mozenda}, iMacros~\cite{iMacros}, 
Visual Web Ripper~\cite{visualwebripper}, Lixto~\cite{lixto}), also known as visual web scraping or visual data extraction, 
is an approach that leverages visual information to extract data from web pages. 
Unlike manual or programmatic methods that rely on HTML structures and coding, visual-based approaches 
utilize computer vision and machine learning techniques to recognize and extract data from the visual elements of web pages.

In visual-based wrapper creation, a user interacts with a tool or framework that provides a visual interface to define the extraction process. 
The user typically highlights the desired data elements on a web page, and the system automatically generates the extraction rules based on the visual cues. 
These extraction rules can include spatial relationships, visual patterns, or templates to identify and extract the relevant data.

One advantage of visual-based wrapper creation is its accessibility to users with limited programming knowledge. 
It allows non-technical users to easily define extraction rules by visually selecting the data elements of interest. 
This makes the process more intuitive and user-friendly, enabling a wider range of individuals to extract data from web pages without extensive coding expertise.
Visual-based wrapper creation also facilitates adaptability to changes in web page layouts. Since it relies on visual patterns rather than HTML structures, 
it can handle variations in web page designs more effectively. When a web page undergoes updates or modifications, 
the visual-based approach can adapt by identifying the visually similar elements for data extraction.

While visual-based wrapper creation offers simplicity and flexibility, it may face challenges with complex or dynamic web pages 
that contain intricate visual elements or require interaction to reveal the desired data. 
Additionally, the accuracy of the extraction heavily relies on the quality of the visual recognition algorithms and the ability to handle variations in visual presentation.

{\bf Comparison with our work}
We have found no published studies papers proposing a method similar to ours.
However, browserless extraction is based on a previous 
paper~\cite{Fayzrakhmanov:2018:BWD:3178876.3186008}.
Our segmentation phase  is inspired by the segmentation 
described in~\cite{DBLP:dblp_journals/tkde/ZhaiL06};
however, to the best of our knowledge, there is no proposal reported in the
literature of discarding rules combined with social extraction and
pointing pattern extraction.
Finally, most of studies attempt to extract all the information from a page 
so that they present the price
of the main entity of interest as well as the price 
of every secondary entity that appears on the page.
They do not include procedures to discriminate between 
the main price and the secondary prices as we do
with our approach.
See, for example,~\cite{DBLP:journals/vldb/FurcheGGSS13}.

\section{Conclusions and Future work}

In this study, we successfully augmented ChatGPT with price extraction functionality 
by implementing our Wextractor architecture. 
This integration was necessitated by ChatGPT's inherent limitation in addressing queries
 requiring browsing capabilities. 
Our approach represents a loosely coupled integration, 
as Wextractor operates independently of ChatGPT's artificial intelligence capabilities.

In addition, enhancements were made to our architecture 
to optimize response time, particularly through improvements in social extraction 
and pointing pattern extraction techniques. These optimizations contribute to a more efficient and timely retrieval of information.

Looking ahead, a critical focus of future research involves extending ChatGPT's capabilities to handle real-time queries requiring browser functionalities. 
This presents a significant challenge that requires careful consideration. One avenue for exploration is the potential reconstruction of Wextractor 
from the ground up, aligning its functionalities more closely 
with the requirements of a large language model such as ChatGPT. 
Alternatively, we must explore methods to integrate the innovative ideas and functionalities 
of Wextractor into the existing ChatGPT framework.

In conclusion, our study represents a significant step forward in augmenting ChatGPT's capabilities for practical applications. 
However, the journey does not end here; rather, it opens up avenues for further research and development. 
By addressing the challenges of real-time browsing capabilities and exploring integration strategies, 
we can continue to enhance the effectiveness and versatility of ChatGPT in addressing 
several user inquiries and tasks.
\bibliography{ppExtraction}{}

\begin{thebibliography}{10}

\bibitem{dataset}
Reproducibility of our price extraction architecture.
\newblock \url{https://acortar.link/96klmh}.
\newblock Accessed: 2024-02-26.

\bibitem{DBLP:dblp_journals/tkde/ChangKGS06}
Chia-Hui Chang, Mohammed Kayed, Moheb~R Girgis, and Khaled~F Shaalan.
\newblock A survey of web information extraction systems.
\newblock {\em IEEE transactions on knowledge and data engineering},
  18(10):1411--1428, 2006.

\bibitem{zipf}
Kenneth~J. Christensen.
\newblock \url{https://cse.usf.edu/~kchriste/tools/genzipf.c}.

\bibitem{Fayzrakhmanov:2018:BWD:3178876.3186008}
Ruslan~R. Fayzrakhmanov, Emanuel Sallinger, Ben Spencer, Tim Furche, and Georg
  Gottlob.
\newblock Browserless web data extraction: Challenges and opportunities.
\newblock In {\em Proceedings of the 2018 World Wide Web Conference}, WWW '18,
  pages 1095--1104, Republic and Canton of Geneva, Switzerland, 2018.
  International World Wide Web Conferences Steering Committee.

\bibitem{ferrara2014web}
Emilio Ferrara, Pasquale De~Meo, Giacomo Fiumara, and Robert Baumgartner.
\newblock Web data extraction, applications and techniques: A survey.
\newblock {\em Knowledge-based systems}, 70:301--323, 2014.

\bibitem{DBLP:journals/vldb/FurcheGGSS13}
Tim Furche, Georg Gottlob, Giovanni Grasso, Christian Schallhart, and
  Andrew~Jon Sellers.
\newblock Oxpath: {A} language for scalable data extraction, automation, and
  crawling on the deep web.
\newblock {\em {VLDB} J.}, 22(1):47--72, 2013.

\bibitem{iMacros}
iMacros.
\newblock \url{http://www.iopus.com/imacros/}.

\bibitem{kung2023performance}
Tiffany~H Kung, Morgan Cheatham, Arielle Medenilla, Czarina Sillos, Lorie
  De~Leon, Camille Elepa{\~n}o, Maria Madriaga, Rimel Aggabao, Giezel
  Diaz-Candido, James Maningo, et~al.
\newblock Performance of chatgpt on usmle: Potential for ai-assisted medical
  education using large language models.
\newblock {\em PLoS digital health}, 2(2):e0000198, 2023.

\bibitem{DBLP:dblp_conf/ijcai/KushmerickWD97}
Nicholas Kushmerick, Daniel~S. Weld, and Robert~B. Doorenbos.
\newblock Wrapper induction for information extraction.
\newblock In {\em IJCAI (1)}, pages 729--737, 1997.

\bibitem{DBLP:dblp_journals/dke/LaenderRS02}
Alberto H.~F. Laender, Berthier~A. Ribeiro-Neto, and Altigran~Soares da~Silva.
\newblock Debye - data extraction by example.
\newblock {\em Data Knowl. Eng.}, 40(2):121--154, 2002.

\bibitem{lixto}
Lixto.
\newblock \url{http://www.lixto.com}.

\bibitem{lloret2020browserless}
Jorge Lloret-Gazo.
\newblock A browserless architecture for extracting web prices.
\newblock In {\em Proceedings of the 35th Annual ACM Symposium on Applied
  Computing}, pages 2193--2200, 2020.

\bibitem{Mozenda}
Mozenda.
\newblock \url{http://www.mozenda.com}.

\bibitem{visualwebripper}
Visual~Web Ripper.
\newblock \url{http://www.visualwebripper.com/}.

\bibitem{DBLP:journals/ftdb/Sarawagi08}
Sunita Sarawagi.
\newblock Information extraction.
\newblock {\em Foundations and Trends in Databases}, 1(3):261--377, 2008.

\bibitem{DBLP:dblp_journals/ml/Soderland99}
Stephen Soderland.
\newblock Learning information extraction rules for semi-structured and free
  text.
\newblock {\em Machine Learning}, 34(1-3):233--272, 1999.

\bibitem{DBLP:dblp_conf/iiwas/StarkaHN13}
Jakub Starka, Irena Holubova, and Martin Necasky.
\newblock Strigil: A framework for data extraction in semi-structured web
  documents.
\newblock In {\em iiWAS}, page 453, 2013.

\bibitem{vasserman2015implementing}
Shoshana Vasserman, Michal Feldman, and Avinatan Hassidim.
\newblock Implementing the wisdom of waze.
\newblock In {\em Twenty-Fourth International Joint Conference on Artificial
  Intelligence}, 2015.

\bibitem{DBLP:dblp_journals/tkde/ZhaiL06}
Yanhong Zhai and Bing Liu.
\newblock Structured data extraction from the web based on partial tree
  alignment.
\newblock {\em IEEE transactions on knowledge and data engineering},
  18(10):1614--1628, 2006.

\end{thebibliography}
\bibliographystyle{plain}

\end{document}